\documentclass[11pt,a4paper]{article}

\pdfoutput=1
\usepackage{jheppub}

\usepackage{amssymb}

\usepackage{braket}

\usepackage{graphicx}
\usepackage{wrapfig,enumerate,slashed}

\usepackage{footmisc}
\usepackage{amsmath}
\usepackage{wasysym} 
\usepackage{color}
\usepackage{comment}
\DeclareMathAlphabet{\mathbold}{U}{zeur}{b}{n}

\usepackage[labelfont=bf,font={sl,small}]{caption}

\renewcommand\[{\left[}
\renewcommand\]{\right]}

\def\beq{\begin{equation}}
\def\eeq{\end{equation}}
\def\[{\begin{equation}}
\def\]{\end{equation}}

\begin{document}
\numberwithin{equation}{section}

\title{Consistency of Higgsplosion in Localizable QFT}

\author{Valentin V. Khoze and Michael Spannowsky}

\affiliation{Institute for Particle Physics Phenomenology, Department of Physics, Durham University, Durham, DH1 3LE, UK}

\emailAdd{valya.khoze@durham.ac.uk}
\emailAdd{michael.spannowsky@durham.ac.uk}

\abstract{We show that large $n$-particle production rates derived in the semiclassical Higgsplosion limit
of scalar field theoretical models with spontaneous symmetry breaking, are consistent 
with general principles of localizable quantum field theory.

\smallskip
\noindent The strict localizability criterium of Jaffe defines quantum fields as
operator-valued distributions acting on test functions that are localized in finite regions of space-time. 
The requirement of finite support of test functions in space-time 
ensures the causality property of QFT. The corresponding localizable fields need not be tempered distributions,
and they fit well into the framework of local quantum field theory.
}

%\date{}

\preprint{IPPP/18/86}

\maketitle

%%%%%%%%%%%%%%%%%%%%%%%%%%%%%%%%%%%%%%%%%%%%%%%%%%%%
%%%%%%%%%%%%%%%%%%%%%%%%%%%%%%%%%%%%%%%%%%%%%%%%%%%%
%%%%%%%%%%%%%%%%%%%%%%%%%%%%%%%%%%%%%%%%%%%%%%%%%%%%

\section{Introduction}

Higgsplosion  \cite{Khoze:2017tjt} is a novel high-energy regime that may be realised in a class of
quantum field theoretical models with microscopic massive scalar fields in (3+1) dimensions~\cite{Khoze:2017ifq}.
This regime is characterised by large transition rates for $few \to many$ particle production processes,
\[
\sqrt{s}\,:\quad X\, \to\,  n\times \phi\,,
\label{eq:Xplosion}
\]
at ultra-high centre of mass
energies $\sqrt{s} \gg m$.
Of particular interest are the 2-particle initial states in a high-energy scattering processes, and the 1-particle initial states
for a very massive or highly virtual particle or a resonance decaying into $n$-particle states. These two types of processes are,
\begin{eqnarray}
{\rm Scattering\, process}: \quad && |X(\sqrt{s})\rangle = |2\rangle \,\,\,\to \,\, |n\rangle \quad \Rightarrow \quad {\rm cross\,\, section} \,\, \sigma_n(\sqrt{s})\,,
\label{eq:2pH}
\\
{\rm Resonance\, decay}: \quad &&|X(\sqrt{s})\rangle = |1^*\rangle \,\to\,\, |n\rangle \quad \Rightarrow \quad {\rm partial\, width} \,\, \Gamma_n(s)\,.
\label{eq:1pH}
\end{eqnarray}
For the 2-particle initial state, the $n$-particle production process \eqref{eq:2pH} is characterised by the cross section $\sigma_n(\sqrt{s})$,
while for the single-particle state of virtuality $p^2=s$ in \eqref{eq:1pH} the relevant quantity is the partial decay width $\Gamma_n (s)$.
Final states contain a large number $n \gtrsim 1/({\rm coupling\, constant}) \gg 1$ of elementary Higgs-like scalar particles $\phi$
of mass $m$.
In particular, if
the partial width of the resonance  $|1^*\rangle$ to decay into $n$ elementary Higgs-like scalars becomes
exponentially large above a certain energy scale $s \gtrsim E_*^2$,
the resonance Higgsplodes; it can be viewed as a composite state of $n$ soft elementary Higgs scalars $\phi$.

Strongest evidence in favour of Higgsplosion comes from the semiclassical calculation in~\cite{Khoze:2018kkz,Khoze:2017ifq}
that is justified in a certain large-$n$ scaling limit. This calculation, results of which we present in section~\ref{sec:2}, is based on
the semiclassical formalism developed earlier in \cite{Son:1995wz} along with the thin-wall technique of~\cite{Gorsky:1993ix}.
Following \cite{Khoze:2018kkz,Khoze:2017ifq,Son:1995wz} we will use a unified description of the Higgsplosion processes 
\eqref{eq:1pH} and \eqref{eq:2pH} in terms of the dimensionless 
quantity ${\cal R}_n (\sqrt{s})$ describing the 
$n$-particle production rate in the semiclassical limit for both processes, neglecting the effect of Higgspersion and the inclusion of appropriate test functions, as discussed in sections~\ref{sec:2} and \ref{sec:3}, we find the proportionality relation
\[
{\cal R}_n (\sqrt{s}) \propto \sigma_n(\sqrt{s}) \propto \Gamma_n (s)\,,
\]
and we will argue that ${\cal R}_n (\sqrt{s}) $ grows exponentially with the energy $\sqrt{s}$ in a Higgsploding theory.
We will explain in section~\ref{sec:2} that ${\cal R}_n (\sqrt{s})$ in the semiclassical approximation is directly related to 
the 1-particle-irreducible (1PI) part $\hat\rho_{n} (s)$ of $n$-particle contribution to the K\"all\'en-Lehmann spectral density,
\[
{\cal R}^{\rm semicl}_n (\sqrt{s}) \propto \hat\rho_{n} (s)\,, \quad {\rm where} \quad
 \hat\rho_{n} (s)  :=\, \int d\Phi_n(m_n^2) \, \delta(s-m_n^2) \, |\langle n|{\cal O}(0)|0\rangle|_{{\rm 1PI}} ^2\,.
\label{eq:rhon}
\]
Here 
${\cal O}(x)$ is a certain local QFT operator, sandwiched on the right hand side of \eqref{eq:rhon}
between the vacuum and the $n$-particle Fock state $|n\rangle$, with $\int d\Phi_n (m_n^2)$ being the integral over the Lorentz-invariant 
$n$-particle phase space and $m_n^2$ is the mass-squared of the state $|n\rangle$.

Hence we have that in a theory with Higgsplosion, the $n$-particle rate, and thus the 1PI spectral density $\hat\rho_{n} (s) $ 
must grow exponentially with the energy,
\[
{\cal R}^{\rm semicl}_n (\sqrt{s}) \, \propto\, \hat\rho_{n} (s) \,\propto \,e^{\,{\rm c}\cdot(\sqrt{s})^\kappa}\,,
\label{eq:Rrhon}
\]
where c is a positive constant and $\kappa > 0$.\footnote{We will explain that the semiclassical expression for the Higgsplosion rate
does not actually predict the value of $\kappa$ in the for us relevant regime of asymptotically high energies $\sqrt{s} \to \infty$ in a given
theory with a fixed value of the (weak) coupling $\lambda$. This is due to the scaling nature of the semiclassical limit where $\sqrt{s} \to \infty$
at the same time as $\lambda \to 0$, so that in order to raise $\sqrt{s}$ one has to lower $\lambda$ in this limit.  Formally, the semiclassical limit 
selects $\kappa= 1$, but quantum corrections allow $\kappa$ to move away from this value.
In particular, values of $\kappa < 1$ required for QFT to be strictly localizable, are entirely consistent with the semiclassical prediction 
for ${\cal R}_n (\sqrt{s}) $. 
}
Equation~\eqref{eq:Rrhon} implies that the 1PI K\"all\'en-Lehmann spectral density, 
\[
\rho(s)_{\,\rm 1PI}   \,=\, \sum_{n} \int dm_n^2\,\,  \hat\rho_n(s)\,,
\]
of the operator ${\cal O}(x)$, grows exponentially with $s$ and 
therefore (at least for a strictly positive $\kappa$) it
cannot be a \emph{tempered} distribution (tempered distributions cannot grow faster than a polynomial).
This may at first appear puzzling, since in the usual Wightman framework of constructive QFT one assumes
that all fields are operator-valued tempered distributions~\cite{Wightman:1956zz,Streater:1989vi}.
However it is also known that the assumption of temperedness is too restrictive.
In fact, a strictly local quantum field theory can be defined in precise mathematical terms
using distributions that are not necessarily tempered and can grow as linear exponentials~\cite{Jaffe:1966an,Jaffe:1967nb}.

\medskip

The aim of this note is to point out that Higgsplosion is perfectly admissible and is not in contradiction with the results or axioms of the suitably 
defined local quantum field theory~\cite{Jaffe:1967nb}.
This is contrary to what was stated recently in Refs.~\cite{Belyaev:2018mtd,Monin:2018cbi} whose considerations in relation
to Higgsplosion relied on 
assuming tempered distributions, in other words the polynomial boundedness of $\rho_n(s)$.
The possibility of more general exponentially growing distributions in a strictly localizable theory was also considered 
in~\cite{Monin:2018cbi}, at least initially, but ultimately dismissed for the case of Higgsplosion.

\medskip

In the next section we will provide a brief summary of the semiclassical origin of the exponential in the Eq.~\eqref{eq:Rrhon},
following \cite{Son:1995wz,Khoze:2018kkz}. Then in section~\ref{sec:3} we will consider the well-known
K\"all\'en-Lehmann expressions for the 2-point Wightman functions of ${\cal O}(x)$ and ${\cal O}(y)$ and their
time-ordered products. In a strictly localizable theory, the fields ${\cal O}(x)$ and ${\cal O}(y)$ are operator-valued distributions 
that are convoluted with the 
test functions of finite support. The test functions are not from the Schwartz space, the corresponding distributions need not be tempered
and admit exponentials in the form appearing in~\eqref{eq:Rrhon} with $0\le \kappa <1$ \cite{Jaffe:1967nb}. 
We will show that Higgsplosion is in fact consistent with the K\"all\'en-Lehmann formulae when we account for the smearing of the operators with 
test functions as is necessary in local QFT. We present our conclusions in section~\ref{sec:4}.

%%%%%%%%%%%%%%%%%%%%%%%%%%%%%%%%%%%%%%%%%%%%%%
\section{Semiclassical Higgsplosion and the K\"all\'en-Lehmann spectral density}
\label{sec:2}
%%%%%%%%%%%%%%%%%%%%%%%%%%%%%%%%%%%%%%%%%%%%%%

A prototype simple model for Higgsplosion is the $\varphi^4$-type real scalar theory in 4 dimensions
with a spontaneously broken $Z_2$ symmetry,
\[
{\cal L} \,=\, \frac{1}{2}\, \partial^\mu \varphi \, \partial_\mu \varphi\, -\,  \frac{\lambda}{4} \left( \varphi^2 - v^2\right)^2
\label{eq:ssb} \,.
\]
The microscopic scalar particles, which play the role of the Higgs bosons, correspond to the excitations of the field $\phi(x) = \varphi(x) -v$
with the bare mass $m_0=\sqrt{2\lambda}v$, and their physical pole mass will be referred to as $m$. 

The probability rate of Higgsplosion ${\cal R}_n (\sqrt{s})$ (cross section in \eqref{eq:2pH} or the partial width in \eqref{eq:1pH}) 
is the integral over the $n$-particle Lorentz-invariant phase space of the amplitude squared,
\[
{\cal R}_n(\sqrt{s}) \,=\,  \int d\Phi_n \, \left| {}^{\rm in}\langle X |n\rangle^{\rm out}_{\sqrt{s}} \right|^2
\,,
\label{eq:RneSfff1}
\]
where ${}^{\rm in}\langle X |$ and $|n\rangle^{\rm out}$ are the initial and final states in the Higgsplosion process \eqref{eq:Xplosion}
and the $\sqrt{s}$ subscript notes that the amplitudes are calculated at 
the centre of mass energy $\sqrt{s}$. Perturbation theory in the regime of Higgsplosion where $n \gtrsim 1/\lambda$,
contains uncontrollable large contributions from powers of $\lambda n \gtrsim 1$ and becomes effectively strongly coupled and cannot be trusted
at any fixed order in $\lambda$. The best currently available non-perturbative technique to compute ${\cal R}_n(\sqrt{s})$ is to rely on a semiclassical 
approximation.
The idea of the semiclassical approach, is that the functional integral representation of the right hand side 
in \eqref{eq:RneSfff1} can be computed in the steepest descent approximation. The large parameter appearing in the exponent of the 
integrals that justifies the steepest descent approach is $n$ -- the particle number in the final state of the Higgsplosion process. All other large
parameters should scale appropriately with $n$ so that \cite{Libanov:1994ug,Son:1995wz,Libanov:1997nt},
\[
n \,\propto\, \sqrt{s}/m \,\propto\, 1/\lambda \, \gg \, 1\,.
\label{eq:lim1}
\]

There is one subtle point in the application of the semiclassical approach to \eqref{eq:RneSfff1}, which is how to describe 
the initial state $|X \rangle$ in the Higgsplosion process. The final state $|n\rangle$ poses no problem as it contains 
$n\sim 1/\lambda \gg 1$ quanta
and is amendable to the semiclassical treatment. The initial state, on the other hand, is not a many-particle state. The resolution advocated in 
\cite{Rubakov:1991fb,Son:1995wz} is to first describe the initial state as a multi-particle state with $c/\lambda$ particles
 in $|X \rangle$ and then take the limit $c\to 0$.
 
Technically, this is achieved by assuming that the initial state is prepared by acting with a certain local operator $\hat{\cal O}(x)$ on the vacuum.
Without loss of generality, by translation invariance one can position this operator at $x=0$,
\begin{equation}
	\ket{X}\,=\,  {\cal O}(0) \ket{0}\,.
\label{eq:opdef1}	
\end{equation}
For carrying out the semiclassical calculation the following choice of the operator is usually made \cite{Son:1995wz},
\[ 
{\cal O}(x) \,=\, j^{-1} \, e^{j \phi(x)}\,,
\label{eq:opdef2}
\]
where $j$ is a constant $j=c/\lambda$. Finally one takes the limit $c\to 0$ (or equivalently $j\to 0$) in the computation of the probability rate
to restrict the initial state $\ket{X}$ in \eqref{eq:opdef1} to the state with the low particle occupation number, as required.

We will assume the operational validity of the prescription in \eqref{eq:opdef1}-\eqref{eq:opdef2} and treat it as a part of the definition
of the semiclassical approach of Son~\cite{Son:1995wz}, on which the calculation in~\cite{Khoze:2017ifq,Khoze:2018kkz}
was based. It is expected that the dependence of the final result for the Higgsplosion rate
on the specific form of the
operator ${\cal O}(x)$ affects only the pre-exponential factor and not the semiclassical exponent of ${\cal R}_n(\sqrt{s})$ in \eqref{eq:RneSfff1}.
The semiclassical exponent itself should not depend on the precise nature of the initial state $X$ as far as it is not a multi-particle state.

%%%%%%%%%%%%%%%%%%%%%%%%%%%%%%%%%%%%%%%%%%%%%%%%%%%%
 \begin{figure*}[t]
\begin{center}
\includegraphics[width=0.85\textwidth]{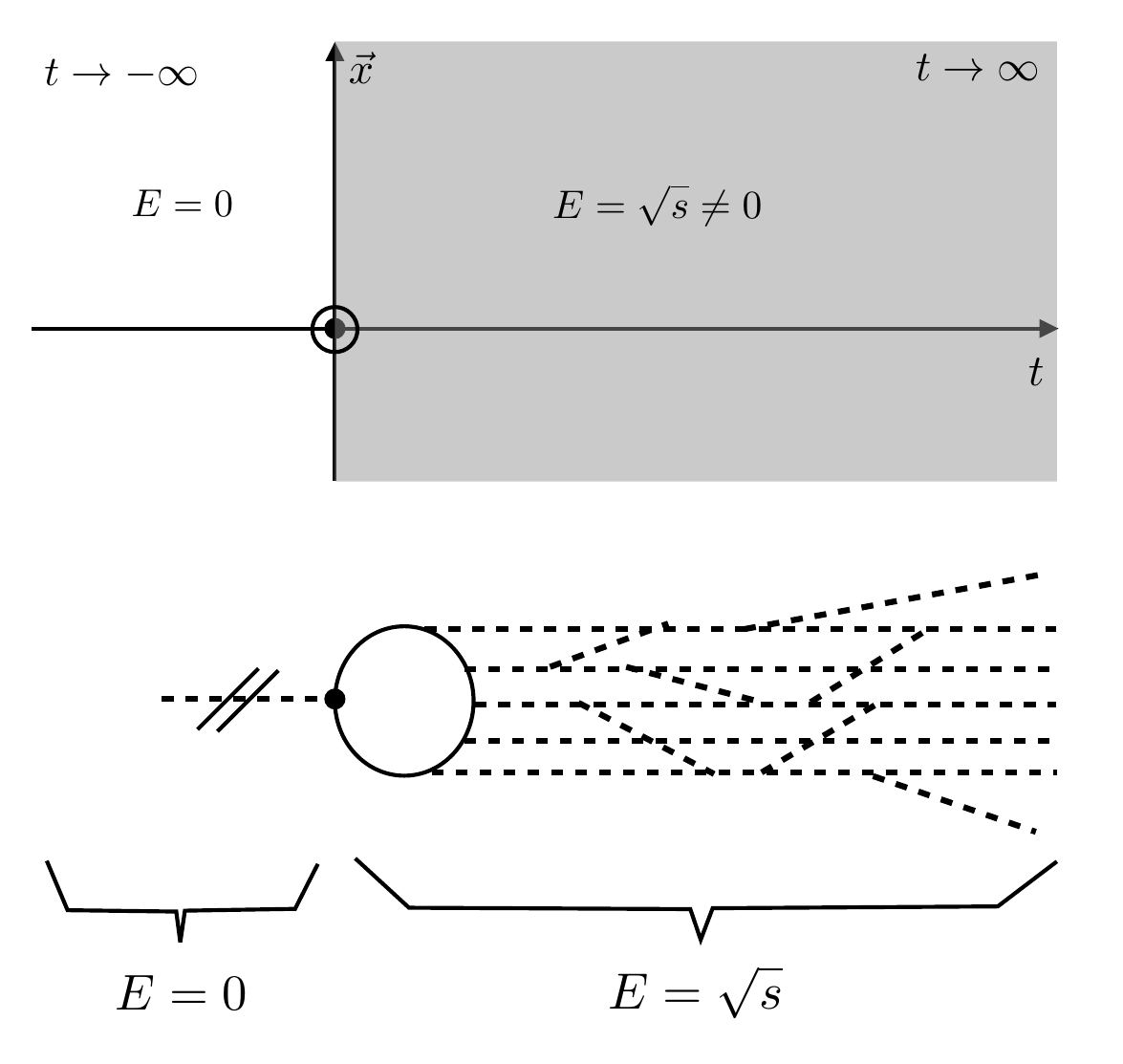}
\end{center}
\vskip-.5cm
\caption{Sketch of a classical field configuration with a single jump in energy at the singular point at the origin $t=0=\vec{x}$ in
Minkowski space. Such configurations can give
 dominant contributions to the 1PI matrix elements but not to the one-particle-reducible ones. 
 The latter would necessarily require more jumps with vanishing energies.}
\label{fig:one}
\end{figure*}
%%%%%%%%%%%%%%%%%%%%%%%%%%%%%%%%%%%%%%%%%%%%%%%%%%%%

%%%%%%%%%%%%%%%%%%%%%%%%%%%%%%%%%%%%%%%%%%%%%%%%%%%%
 \begin{figure*}[t]
\begin{center}
\includegraphics[width=0.85\textwidth]{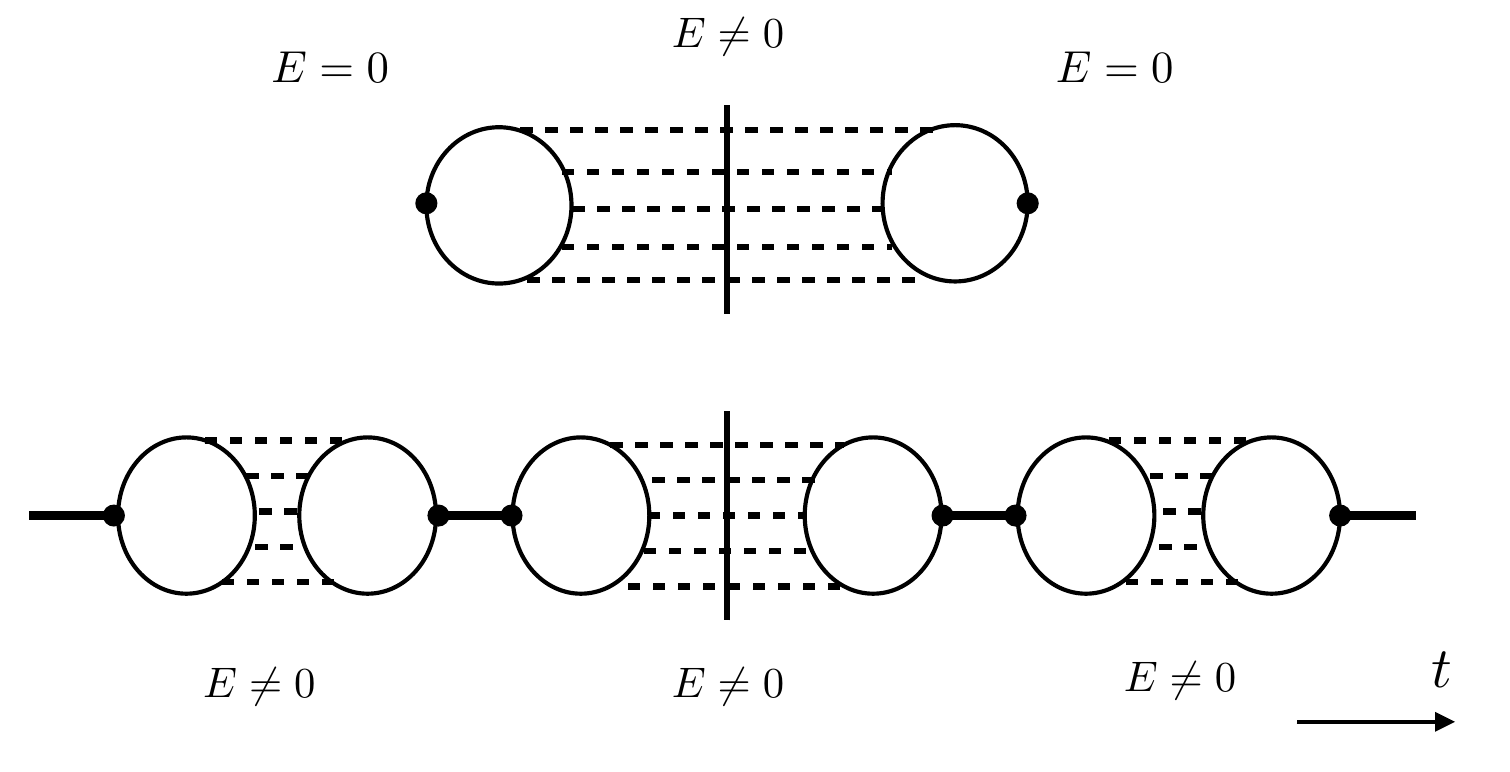}
\end{center}
\vskip-.5cm
\caption{One-particle-reducible contributions require saddle-point configurations with multiple singular points in Minkowski space.}
\label{fig:two}
\end{figure*}
%%%%%%%%%%%%%%%%%%%%%%%%%%%%%%%%%%%%%%%%%%%%%%%%%%%%

The amplitude ${\cal M}_{1^*\to n}$ for the $X(\sqrt{s})$-resonance decay $|X\rangle \to |n\rangle $ is given by the
1-particle-irreducible matrix element,
\[ 
{\cal M}_{1^*\to n}^\dagger\,=\, {}^{\rm in}\langle X |n\rangle^{\rm out}_{\quad \rm 1PI}\,=\, 
 \langle 0|{\cal O}^\dagger(0)\,{\cal S}^\dagger|n\rangle_{\rm 1PI}\,,
 \label{eq:ampl1PI}
 \]
where ${\cal S}$ is the $S$-matrix and the outgoing $n$-state has the COM energy $\sqrt{s}$. The 1-incoming and $n$-outgoing 
external lines of the matrix element on the right hand side are LSZ-amputated. In the semiclassical approach one evaluates the path
integral representation of $\langle 0|{\cal O}^\dagger(0)\,{\cal S}^\dagger|n\rangle_{\rm 1PI}$ in the saddle-point approximation, expanding 
around a classical solution which satisfies the appropriate boundary conditions at $t \to \pm \infty$. As explained in 
Refs.~\cite{Son:1995wz,Khoze:2018mey},
these are such that at $t\to -\infty$ the solution contains only the positive frequency components,
while at $t\to +\infty$ it has both the positive and the negative frequency components.
 As the result the energy $E$ of the solution is vanishing at all
 $t$ in the interval $-\infty <t<0$, and is non-vanishing and equal to $\sqrt{s}$ for $0<t<+\infty$. The solution is singular at the origin,
 $x_\mu=0$, where the operator ${\cal O}$ in \eqref{eq:ampl1PI} is located, and the presence of this singularity explains the jump 
 in the energy of the classical solution from $E=0$ to $E=\sqrt{s}$ when time passes from $t<0$ to $t>0$.
 
 Such classical saddle-point solutions are depicted schematically in Minkowski space $(\vec{x},t)$ in Fig.~\ref{fig:one}.
 It should be clear from this figure that such field configurations with a single jump in energy from $0$ to $\sqrt{s}$ at the unique
 singular point $x=0$ can correspond only to one-particle irreducible contributions to the matrix element. More precisely, any 
 one-particle-reducible contributions would require field configurations changing their classical energy from $0$ to $E$ at a point $t_1$,
 then from $E$ back to $0$ at a point $t_2>t_1$, then from $0$ to $\sqrt{s}$ at a point $t_3>t_2$. This is depicted in Fig.~\ref{fig:two}.
 Hence we conclude that the simple
 saddle-point solutions that have a single energy jump at a single singularity point in Minkowski space -- which are the saddle-points
 considered in the semiclassical approach -- approximate the one-particle-irreducible matrix elements, as indicated by the 1PI 
 subscript on the right hand side of \eqref{eq:ampl1PI}.

It then follows that the partial decay width of the resonance $X$ with the virtuality $s$ expression in \eqref{eq:RneSfff1} can be written as,
\[
\Gamma_n(s)\,\propto\, {\cal R}_n^{\rm semicl}(\sqrt{s}) \,=\,  \int d\Phi_{n}(s)\, \, \langle 0|{\cal O}^\dagger(0)\,{\cal S}^\dagger|n\rangle_{\rm 1PI} \, \langle n| {\cal S} \,{\cal O}(0) |0\rangle_{\rm 1PI}
\,.
\label{eq:RneSfff}
\]
The phase space volume element  $d\Phi_{n}(s)$ 
 in \eqref{eq:RneSfff} is the standard $n$-particle bosonic Lorentz-invariant phase space,
\[ 
\int d\Phi_{n} (p^2)
\,\,= \,\, \frac{1}{n!}\,
 \prod_{j=1}^n \int \frac{d^3 k_j}{(2\pi)^3\, 2 k_j^0} \,\, (2\pi)^4 \delta^{(4)}(p-\sum_{j=1}^n k_j)\,,
 \label{eq:dPhi}
 \]
 computed at $p^2=s$, where $p_\mu$ is the total momentum in the reaction. 
% The matrix elements include the S-matrix ${\cal S}$,  and its Hermitian conjugate ${\cal S}^\dagger$, so that it is no longer necessary to distinguish between the in- and out-states in \eqref{eq:RneSfff}.
 
 \medskip
 
 Anticipating the discussion of admissibility of Higgsplosion in the formal local QFT framework in the next section,
it is worthwhile to note here that quantum fields are not operators acting on the Hilbert space of states, but operator-valued distributions
 \cite{Wightman:1956zz,Streater:1989vi,Bogolyubov:1990kw}.  This leads to a straightforward modification of the semiclassical
 prescription \eqref{eq:opdef1}-\eqref{eq:opdef2}  for the definition of the initial state $\ket{X}$, which proceeds as follows.
 Since any field that is sharply defined at a point $x$, is a distribution, to define an operator one has to smear the field with a test function
 that belongs to an appropriate set of well-behaved smooth and rapidly decreasing functions.
This implies that  ${\cal O}(x)$ in \eqref{eq:opdef2} should be averaged with a test function $g(x)$.
The operator localised in the vicinity of a point $x$ is then,
\[
{\cal O}_g (x)\,=\, \int d^4 x' \, g(x'-x)\, {\cal O}(x')\,,
\label{eq:2.8}
\]
and the prescription \eqref{eq:opdef1} for defining the initial state is refined using,
\[
	\ket{X}\,=\,  {\cal O}_g(0) \ket{0}\,=\, \int d^4 x' \, g(x')\, {\cal O}(x') \, \ket{0}
	\,.
\label{eq:opdef11}	
\]
This gives a well-defined state in the Hilbert space. For the rest of this section we will temporarily ignore
the averaging of the operators with the test functions. Their effect is easily recovered from the distribution-valued 
rate ${\cal R}_n^{\rm semicl}(\sqrt{s})$ that we will now compute.

\medskip

The semiclassical rate ${\cal R}_n^{\rm semicl}(E)$ for the Higgsplosion process $X \to n\times \phi$ in \eqref{eq:RneSfff}  
was computed in~\cite{Khoze:2017ifq,Khoze:2018kkz} to exponential accuracy in the scalar theory \eqref{eq:ssb}. 
In general the validity of the semiclassical approach requires working in the steepest descent limit {\it cf.}~\eqref{eq:lim1},
\[ \lambda \to 0\,, \,\,\, n\to \infty\,,  \,\,\, \sqrt{s}/m\to \infty\,, \,\,\, {\rm with}\,\,\,
\lambda n = {\rm fixed}\,, \,\,\, \varepsilon :=\, \frac{\sqrt{s}\,-\, nm}{nm}={\rm fixed} \,,
\label{eq:limit}
\]
where the parameter $\varepsilon$ is the kinetic energy in the final state per particle per mass.

On general grounds, the semiclassical prediction for the rate in the double-scaling limit \eqref{eq:limit} should be of the 
form~\cite{Libanov:1994ug,Son:1995wz},
\[
{\cal R}_n^{\rm semicl}(\sqrt{s}) \,=\,  e^{ \,n\, F(\lambda n, \varepsilon)}
\,,
\label{eq:Rnhg}
\]
where $F(\lambda n, \varepsilon)$ is some function of two arguments, both of which are kept fixed in the semiclassical limit \eqref{eq:limit}.
At small values of $\lambda n$, the function $F(\lambda n, \varepsilon)$ is known and is negative-valued, hence there is no Higgsplosion 
at relatively low multiplicities, and the rate in \eqref{eq:Rnhg} is exponentially suppressed\footnote{In the regime $\lambda n \ll 1$,
ordinary perturbation theory is a valid, and the semiclassical
computation carried out in \cite{Son:1995wz} correctly reproduced the previously known perturbative 
results~\cite{Brown:1992ay,Voloshin:1992nu,Smith:1992rq,Libanov:1994ug}.}.
The regime where the function $F(\lambda n, \varepsilon)$ can potentially become positive 
and result in ${\cal R}_n^{\rm semicl}(\sqrt{s})$ growing exponentially with $n\sim \sqrt{s}/m$,
would only be possible at sufficiently large values of $\lambda n$.
Fortunately, the semiclassical approach is equally applicable in this 
{\it non-perturbative} regime where we take,
\[ 
\lambda n = {\rm fixed} \gg1 \,, \quad \varepsilon ={\rm fixed} \ll 1 \,.
\label{eq:limit2}
\]
This calculation was carried out in Refs.~\cite{Khoze:2017ifq,Khoze:2018kkz}, with the result given by
\[
{\cal R}_n^{\rm semicl}(\sqrt{s})
\,= \, 
\exp \left[ n\, \left( 
\log \frac{\lambda n}{4} \,+\, 0.85\, \sqrt{\lambda n}\,+\,\frac{1}{2}\,+\,\frac{3}{2}\log \frac{\varepsilon}{3\pi} 
\, -\,\frac{25}{12}\,\varepsilon 
\right)\right]\,,
\label{eq:Rnp2}
\]
which corresponds to
\[
F(\lambda n, \varepsilon)\,=\, 
\log \frac{\lambda n}{4} \,+\, 0.85\, \sqrt{\lambda n}\,+\,\frac{1}{2}\,+\,\frac{3}{2}\log \frac{\varepsilon}{3\pi} 
\, -\,\frac{25}{12}\,\varepsilon \,,
\label{eq:Rnp22}
\]
in this limit.
The expression \eqref{eq:Rnp2} was derived in the near-threshold limit where final state particles are non-relativistic so that $\varepsilon$ 
is treated as a fixed number much smaller than one. 
The overall energy  and the final state multiplicity are related linearly via
$\sqrt{s}/m \,=\, (1 + \varepsilon) \, n \simeq n \gg 1 $.
Clearly, for any small fixed value of $\varepsilon$ one can choose a sufficiently large value of $\lambda n$, such that the function 
$F(\lambda n, \varepsilon)$ in  \eqref{eq:Rnp22} is positive. (See the discussion in~\cite{Khoze:2017ifq} for more detail.)

The semiclassical results \eqref{eq:Rnp2}-\eqref{eq:Rnp22} imply that at sufficiently large particle multiplicities, the 
expression ${\cal R}_n^{\rm semicl}(\sqrt{s})$ grows exponentially with $n$ and consequentially with the energy $\sqrt{s}$.

\medskip

We now recall from our earlier discussion
that the expressions for ${\cal R}_n^{\rm semicl}(\sqrt{s}) $ in \eqref{eq:Rnp2} and \eqref{eq:Rnp22kappa} are
in fact distribution-valued functions. To obtain the proper $n$-particle production rate one needs to account for the operator-smearing 
effect in the definition of the initial state in \eqref{eq:opdef11}.
The result of this is that the Higgsplosion rate becomes $|\tilde{g} (\sqrt{s})|^2 \, {\cal R}_n^{\rm semicl}(\sqrt{s})$ where $\tilde{g}(p)$ is the Fourier transform
of the test function $g(x')$ in space-time to the to momentum space. This implies that the Higgsplosion rate can be written in the form,
\[
R_g^{\rm semicl}(n,\sqrt{s})\,=\, |\tilde{g} (\sqrt{s})|^2 \,\, {\cal R}_n^{\rm semicl}(\sqrt{s})\, =\, 
|\tilde{g} (\sqrt{s})|^2 \, e^{ \,n\, F(\lambda n, \varepsilon)}
\,,
\label{eq:ggRnhg}
\]
by dressing the leading order semiclassical result ${\cal R}_n^{\rm semicl}(\sqrt{s})$ with the smearing function $|\tilde{g} (\sqrt{s})|^2$. 
This smearing will also ensure an
acceptable behaviour of the physical production rate at asymptotically high centre of mass energies, in accordance with unitarity.

\medskip

An important question to answer for establishing whether Higgsplosion contradicts or is in tension with fundamental principles of local
QFT is how fast the
expression for ${\cal R}_n(\sqrt{s})$, grows with $\sqrt{s}$ as a distribution\footnote{ i.e. ignoring the effect of averaging with test functions.}
at asymptotically high energies in a given theory with fixed value of $\lambda$. 
As we already mentioned in the Introduction, the multi-particle production rate ${\cal R}_n^{\rm semicl}(\sqrt{s})$ is closely related to
the $n$-particle contribution $\hat\rho_n (s)$ to the K\"all\'en-Lehmann spectral density,
as can be seen from their defining expressions,
\begin{eqnarray}
{\cal R}_n^{\rm semicl} (\sqrt{s}) &=& \int d\Phi_{n}(s)\, \, \langle 0|{\cal O}^\dagger(0)\,{\cal S}^\dagger|n\rangle_{\rm 1PI} 
\, \langle n| {\cal S} \,{\cal O}(0) |0\rangle_{\rm 1PI}
\,, \label{Rnsform}\\
\hat\rho_n(s) &=& \int d\Phi_n(m_n^2)\,\, \delta(s-m_n^2) \, \langle 0|{\cal O}^\dagger(0)|n\rangle_{\rm 1PI} 
\, \langle n| {\cal O}(0) |0\rangle_{\rm 1PI}\,.
\label{rhonsform}
\end{eqnarray}
Without loss of generality, we can parameterise the exponential growth of the Higgsplosion rate with the energy in the form
\[
{\cal R}_n^{\rm semicl}(\sqrt{s}) \,\sim \, 
 \exp\left[c\, (\sqrt{s})^{\kappa}\right],
\label{eq:Rnp22kappa}
\]
for a positive constant $c$ and some positive power $\kappa$, and we expect that the same exponential 
characterises the behaviour of the spectral density,
\[
\rho(s)_{\rm 1PI}  \,=\, \sum_{n=1}^\infty \int dm_n^2\, \int d\Phi_n(m_n^2)\,\, \delta(s-m_n^2) \, |\langle n| {\cal O}(0) |0\rangle|^2
\,\sim\,  \exp\left[c\, (\sqrt{s})^{\kappa}\right].
\label{eq:spnp22kappa}
\]
How fast the spectral density is allowed to grow in the asymptotic high-energy limit in a given theory, i.e. in the limit where
\[
\lambda ={\rm fixed} \,, \quad {\rm and} \quad \sqrt{s} \sim m n \to \infty\,,
\label{eq:limphys}
\]
determines what kind of distribution it is and which distributions are allowed in the field theoretical framework.

\medskip

Our task now is to determine if Higgsplosion can predict the range for the $\kappa$ parameter and thus determine what type 
of distributions the spectral density belongs to and if this type is admissible in a local QFT framework.
Does the semiclassical Higgsplosion rate \eqref{eq:Rnp2} fix the parameter $\kappa$ in the equation \eqref{eq:Rnp22kappa}?
We will now explain that it does not. 

\medskip 

Starting with the expression \eqref{eq:Rnp2}, one could naively expect that the contribution 
${\cal R}_n(\sqrt{s})\sim e^{\,n\sqrt{\lambda n}}$ in the limit \eqref{eq:limit} gives the high-energy asymptotics
${\cal R}_n(\sqrt{s})\sim e^{ \, (\sqrt{s})^{1.5}}$.
Note, however, that promoting $n\sqrt{\lambda n}$ to $(\sqrt{s})^{1.5}$ is at odds with the semiclassical limit \eqref{eq:limit}, which requires 
that $\sqrt{\lambda n}$ is held fixed (rather than scales as $\sqrt{s}/m$) as  $\sqrt{s} \to \infty$.
In practice, the expression on the right hand side of \eqref{eq:Rnp22} can (and in general will) 
receive power series corrections of the form 
$\lambda^p (\lambda n)^q$,  with positive $p$ and $q$. 
Such contributions are not accounted for in the leading order semiclassical expressions.\footnote{They vanish in the semiclassical limit~\eqref{eq:limit}, 
where $\lambda \to 0$
and  $\lambda n=$~fixed.} 
But in the `physical' high-energy limit \eqref{eq:limphys} where we are considering the high-energy asymptotic behaviour within the same theory
so that $\lambda$ is held fixed (possibly modulo slow logarithmic running),
these corrections cannot be ignored. For $q > 1/2$ they dominate over the $\sqrt{\lambda n}$ term in $F(\lambda n, \varepsilon)$,
thus invalidating the $\kappa = 1.5$ assumption.

It is more prudent to treat $F(\lambda n, \varepsilon)$ as a constant -- at least to be consistent with the semiclassical limit. 
But even then, it would be incorrect to
claim that  the rate grows precisely as the {\it linear} exponential with $\kappa =1$, i.e. that
$\log {\cal R}_n(\sqrt{s}) \sim n^1 \sim (\sqrt{s})^1$. Clearly, in the semiclassical limit one cannot distinguish between 
$n^{1}$ and $n^{1-\lambda}$ because the quantity $\lambda \log n $ distinguishing the two vanishes in the limit \eqref{eq:limit},
$\lambda \log n \to 0$.

To summarise the discussion above,
we conclude that the semiclassical expression for the Higgsplosion rate
does not predict the value of $\kappa$ in the relevant for us regime \eqref{eq:limphys}, where we compare the asymptotic high-energy 
behaviour within the same theory,
i.e. the theory with a {\it fixed} coupling $\lambda\ll 1$. 
Formally, the semiclassical limit 
selects $\kappa= 1$, but quantum corrections to the leading order scaling expressions allow $\kappa$ to freely deviate from this value.
For concreteness, in most of what follows we will assume that
\[
0< \kappa < 1\,,
\] 
which is entirely consistent with the semiclassical prediction 
for ${\cal R}_n (\sqrt{s})$ and, as we will explain the following section, corresponds to the case of strictly localizable QFTs.\footnote{The case
of $\kappa=1$ is the quasi-localizable case and $\kappa >1$ gives a QFT framework that cannot be localised in space-time.}
Our aim is not to prove that Higgsplosion implies $\kappa < 1$, but to investigate whether it leads to any inconsistencies 
with a reasonable local field theory setup, and if it does, what is the price to pay for having Higgsplosion.
For the theory to be local we need $\kappa <1$, and we have argued that this regime does not contradict anything we know about
Higgsplosion from general principles.

%%%%%%%%%%%%%%%%%%%
\section{Strictly localizable fields and the self-consistency of Higgsplosion}
\label{sec:3}
%%%%%%%%%%%%%%%%%%%

In the axiomatic formulation \cite{Streater:1989vi, Bogolyubov:1990kw}, the characterisation of a QFT model and all its properties are encoded 
in the Wightman functions,
of local operators $O(x)$
\[
W_n (x_1,\ldots, x_n)\,=\, \langle 0 | O(x_1)\ldots O(x_n)|0\rangle\,.
\label{eq:Whit}
\]
The `operators' $O (x)$
and their Wightman functions \eqref{eq:Whit} are understood in the sense of distributions.

Operator-valued distributions are linear functionals that map a set of test functions into operators acting on the Hilbert space of states.
If $O(x)$ is an operator-valued distribution in the coordinate space and $g(x)$ is a test function, the linear map is
\[
O_g \,=\, \int d^4 x\, O(x)\, g(x)\,,
\label{eq:distint}
\]
where $O_g$ is an operator that acts on the Hilbert space.
We will consider two classes of test functions and distributions. In the first case we will require that the test function $g(x)$ is 
1) smooth (infinitely differentiable), and
2) has compact support in the 4-dimensional space. We will call the space of such test functions $D$.
The distributions $O(x)$ belong to the dual space $D'$ which is defined by requiring that the integral in \eqref{eq:distint} is finite. 
We  require that the Fourier transform of distributions exists and $\int d^4 x\, O(x)\, g(x)= \int d^4 p\, \tilde{O}(p)\, \tilde{g}(-p)$.
The distributions $O(x) \in D'$ will be called {\it strictly localizable}, reflecting the property of the test functions having support on finite, i.e. compact,
regions in the coordinate space. In momentum space the distributions $\tilde{O}(p)$ will grow no faster than
\[
\tilde{O}(p) \sim\, P_N(|p|)~e^{\,c\, |p|^{\kappa}}
\label{eq:kappa}
\]
with $0\le \kappa <1$ and $P_N$ represents a polynomial of order N for any finite N.

The other class of the distributions we are interested in are {\it tempered distributions}.
Their test functions $g(x)$ belong to the Schwartz space S. They are 1) infinitely differentiable functions which 2) are rapidly 
decreasing at $|x|\to \infty$
along with any number of partial derivatives, i.e. for $g(x) \in S$ one has $\lim_{|x| \to \infty} x^n g(x) \to 0~\forall~n \in \mathbb{N}$.
Thus the test functions from the Schwartz space are peaked and rapidly falling, but are not required to have finite support.
The corresponding distributions belong to the dual space $O(x) \in S'$ and are called tempered distributions. Fourier transform 
of a tempered distribution is a tempered distribution. In both coordinate and momentum representations, the growth of tempered distributions is
bound by a fixed order polynomial.

The test functions spaces satisfy $ D \subset S$ since functions with finite support form a subspace of the functions which are rapidly decreasing
at large $x$. On the other hand, the corresponding distribution spaces are ordered in the opposite way, thanks to \eqref{eq:distint}
 $S' \subset D'$. All tempered distributions are strictly localizable and correspond to a special case $\kappa=0$. 
 
\medskip

A useful tool for understanding which types of operator-valued distributions can be allowed in QFT is the 
K\"all\'en-Lehmann spectral decomposition formula (see Eq.~\eqref{eq:spWa} below) for the 2-point Wightman function, 
\[
W (x, y)\,=\, \langle 0 | O^\dagger (x) \,O(y)|0\rangle\,.
\label{eq:Whit2}
\]
The spectral decomposition is derived by inserting 
the sum 
% $\sum_\alpha \,:=\, \sum_n \,\int dm_n^2\,\int d\Phi_n(m_n^2)$ 
over a complete set of states,
\[
1\,=\, |0 \rangle \langle 0| \,+\, 
 \sum_\alpha\,  \int\frac{d^3 p}{(2\pi)^3} \frac{1}{2 E_{\bf p}(\alpha)} \,\,  |\alpha_{\bf p} \rangle \langle \alpha_{\bf p}|\,,
\label{eq3}
\]
between the two operators on the right hand side of \eqref{eq:Whit2}.
Here $|\alpha_{\bf p}\rangle$ are the relativistically normalised
$n$-particle states, Lorentz-boosted to the frame with the total 3-momentum ${\bf p}$ and the energy 
$E_{\bf p}(\alpha)=\,\sqrt{|{\bf p}|^2 +m_\alpha^2}$, where $m_\alpha^2$ is the invariant mass of the $\alpha$-state.

Our notation for the summation over the multi-particle states in \eqref{eq3}
is as follows,
\begin{eqnarray}
\sum_\alpha\,  \int\frac{d^3 p}{(2\pi)^3} \frac{1}{2 E_{\bf p}(\alpha)} \,\,  |\alpha_{\bf p} \rangle \langle \alpha_{\bf p}|
&:=&  \int\frac{d^3 p}{(2\pi)^3} \frac{1}{2 E_{\bf p}} \,\,  |1_{\bf p} \rangle \langle 1_{\bf p}|
\nonumber \\
&+&
\sum_{n=2}^\infty\, \int_{(nm)^2}^\infty d m_n^2  \int\frac{d^3 p}{(2\pi)^3} \frac{1}{2 E_{\bf p}(m_n^2)} \,\int  d\Phi_n(m_n^2)\,\,
 |n_{\bf p} \rangle \langle n_{\bf p}|\,,
\nonumber
\end{eqnarray}
where $E_{\bf p}(\alpha)\equiv\, E_{\bf p}(m_n^2)=\,\sqrt{|{\bf p}|^2 +m_n^2}$ and we use the notation $m_\alpha^2 \equiv m_n^2$ 
to denote the invariant mass of the corresponding state $ |\alpha_{\bf p} \rangle \equiv |n_{\bf p} \rangle $.

The Poincar{\'e} invariance 
implies,
\begin{eqnarray}
O(x) = e^{i P^\mu x_\mu} \, O(0) \, e^{-i P^\mu x_\mu} \,\,\Rightarrow
 &&\langle 0| O^\dagger(x)|\alpha_{\bf p} \rangle \,=\, e^{-ip\cdot x} \langle 0| O^\dagger(0)|\alpha \rangle \,, 
 \nonumber \\
&&\,\,\,  \langle \alpha_{\bf p}|O(y)| 0\rangle \,=\, e^{ip\cdot y}\, \langle \alpha|O(0)| 0\rangle,
\label{eq5}
\end{eqnarray}
and it follows that,
\begin{eqnarray}
W(x,y) &=&
 \sum_\alpha\,  \int\frac{d^3 p}{(2\pi)^3} \frac{1}{2 E_{\bf p}(\alpha)} \, e^{-i p\cdot(x-y)}\, |\langle \alpha | O(0)|0 \rangle |^2
  \label{ewq:sp1} \\
 &=&
 \sum_\alpha\, D^{(0)}(x-y; m_\alpha^2) \,\, |\langle \alpha| O(0)|0\rangle |^2 \,.
\label{eq6}
\end{eqnarray}
The  expression $D^{(0)}(x-y; m_\alpha^2)$ introduced in \eqref{eq6} is the 2-point function of free fields,
\[ 
D^{(0)}(x-y; m^2) \,=\, \langle 0| \phi(x) \phi(y)|0 \rangle_{\rm free}
 \,=\,
\int\frac{d^3 p}{(2\pi)^3} \frac{1}{2 E_{\bf p}} \, e^{-i p\cdot(x-y)}\,,
\label{eq:D0}
\]
with the mass-squared 
parameter $m^2$ replaced by $m_\alpha^2$.
Finally, inserting $1= \int ds \, \delta(s-m_\alpha^2)$ on the right of \eqref{eq6} and exchanging the order of the
summation over the complete set and the integral over $s$, we obtain the K\"all\'en-Lehmann spectral decomposition of $W(x,y)$:
\[
W(x,y) \,=\, 
 \langle 0 | O(x) O^\dagger(y)|0\rangle \,=\, 
\int_0^{\infty} ds \, \rho(s)\, D^{(0)}(x-y; s) \,,
\label{eq:spWa}
\]
where $\rho(s)$,
\[
\rho(s) \,=\, \sum_{\alpha}\, \delta(s - m_\alpha^2)\, |\langle \alpha| O(0)|0 \rangle |^2\,.
\label{eq:rhodef}
\] 
is the spectral density. {\it Unitarity} implies that $\rho(s)\ge 0$ for all $s\ge 0$ and {\it stability} that there are no tachyons and $\rho(s)$ only has support for $s\ge 0$.
Both these properties 
follow from the defining expression \eqref{eq:rhodef}.

\medskip

To make a connection with the semiclassical Higgsplosion rate of the previous section, we can now isolate the one-particle-irreducible
part the spectral density and correspondingly of the Wightman 
function by writing,
\[
W(x,y)_{\rm 1PI} \,=\, 
 \langle 0 | O(x) O^\dagger(y)|0\rangle_{\rm 1PI} \,=\, 
\int_0^{\infty} ds \, \rho(s)_{\,\rm 1PI}\, D^{(0)}(x-y; s) \,.
\label{eq:spWa1PI}
\]
Note that if the operators ${\cal O}$ and $\phi$ do not mix, the overlap with the 1-particle state is automatically zero,
$\langle 0|{\cal O}(x) |1\rangle \,=\, 0$. In this case the spectral density in \eqref{eq:rhodef}
and the Wightman function in \eqref{eq:spWa} are automatically 
one-particle irreducible,  $\rho(s)\,=\, \rho(s)_{\rm 1PI}$ and
$W(x,y)\,=\, W(x,y)_{\rm 1PI}$.

\medskip

The central question for us is how fast can the distribution $\rho(s)_{\rm 1PI}$ be allowed to grow at $s\to \infty$ for the integral 
$\int_0^\infty ds$ on the right hand side of \eqref{eq:spWa1PI} to be finite, so that the Wightman function $W(x,y)_{\rm 1PI}$ 
defined by \eqref{eq:spWa1PI} even exists in the coordinate space.

\medskip

A \emph{strictly localizable} field $O(x)$ is one for which the spectral density integral is finite and the Wightman function
$W(x,y)_{\rm 1PI}$ in
\eqref{eq:spWa1PI} is well-defined for $x\neq y$. Only the vicinity of $x=y$ needs to be avoided and this is achieved by 
averaging or smearing the distribution-valued operators with test functions of \emph{compact} support, as in \eqref{eq:2.8}.
Jaffe~\cite{Jaffe:1967nb} proved that the requirement of 
strict localizability implies that 
the distributions cannot grow faster than,  
\[
\rho(s)_{\rm 1PI} \sim\, \exp\left[c\, (\sqrt{s})^{\kappa}\right], \quad{\rm where} \quad 0\le \kappa <1\,.
\label{eq:strictloc}
\]
To see that this is indeed the case,
we can use the asymptotic properties of the free-theory function $D^{(0)}(x-y; m^2)$ in \eqref{eq:D0} at $m^2=s$,
\begin{eqnarray}
&D^{(0)}(x-y; s)& \approx \frac{(2\sqrt{s})^{1/2}}{(4\pi|x-y|)^{3/2}}\,e^{-\sqrt{s}|x-y|}\,, \quad (x-y)^2<0\,,\quad \sqrt{s}|x-y|\gg 1,
\qquad \label{eq:Wf1}
\\
&D^{(0)}(x-y; s)& \approx \frac{(2\sqrt{s})^{1/2}}{(4\pi|x-y|)^{3/2}}\,e^{-i\sqrt{s}|x-y|}\,, \quad (x-y)^2>0\,,\quad \sqrt{s}|x-y|\gg 1,
\label{eq:Wf2}
\end{eqnarray}
and substitute these expressions into the spectral decomposition formula \eqref{eq:spWa1PI}.
Because of the linear exponential cut-off provided by the expressions in \eqref{eq:Wf1} and \eqref{eq:Wf2}
we see that
the integral over $s$ converges for all distributions $\rho(s)$ that grow slower than a linear exponential, in agreement with what
is indicated in \eqref{eq:strictloc}.
The distributions that grow as 
$e^{\,c\, (\sqrt{p^2})^{\kappa}}$  in momentum space require test functions 
that fall off sufficiently fast, i.e. not slower than $e^{\,-c'\, (\sqrt{p^2})^{\kappa}}$. 
It is known that there exist smooth test functions with compact support in 
space-time, such that their Fourier transforms to momentum space are of this form
when 
$\kappa<1$ \cite{Meiman,Jaffe:1967nb}. 
The requirement of compact support of test functions in space-time 
ensures the causality property of QFT. The operators,
\[
O_g \,=\, \int d^4 x\, O(x)\, g(x) \,, \quad {\rm and} \quad
O_f \,=\, \int d^4 x\, O(x)\, f(x)\,,
\]
commute whenever test functions $g(x)$ and $f(x)$ are localised on spacelike separated regions.

In comparison, for the distributions with $\kappa \ge1$ the test functions with compact support in coordinate space do not exist.
Test functions in momentum space, $\tilde{g}(p)$, that are infinitely differentiable and which are bounded by
 $\tilde{g}(p)< C\,e^{\,-c'\, (\sqrt{p^2})^{\kappa}}$,
have Fourier transforms which are not more localised in space-time than $g(x) \approx e^{\,-c\, |x-a|^{\frac{\kappa}{\kappa-1}}}$.
These test functions are non-vanishing over the entire coordinate space, and the system cannot be localised in space-time
\cite{Gelfand,Keltner:2015xda}.

Following this line of reasoning we conclude that Higgsplosion is consistent with strictly localizable distributions, e.g. of the form Eq.~\eqref{eq:strictloc}, which ensure that the Wightman functions of Eq.~\eqref{eq:spWa1PI} are well-defined. 
The corresponding QFT admits a local interpretation \cite{Jaffe:1967nb} and results in a spectral density that respects unitarity and stability.  
The expression on the right hand side of~\eqref{eq:spWa1PI} is well-defined for all values of the arguments, except at
$(x-y)^2 = 0$
where the Wightman function  becomes singular.
This short-distance singularity can however be avoided by smearing the operators with test functions, as in \eqref{eq:distint}.

\bigskip
\medskip

Why is it then often assumed that the spectral density should be further restricted to a tempered distribution? 
The growth of tempered distributions is bounded by a finite-order polynomial rather than exponential, so they correspond to
a particular subspace of strictly localizable distributions with $\kappa=0$ in Eq.~\eqref{eq:kappa}.

The attractiveness of tempered distributions is motivated by the second K\"all\'en-Lehmann spectral formula, Eq.~\eqref{eq:disp1}, 
i.e. for the time-ordered products.
From the definition of time-ordering,
\[
\langle 0 | T(O(x) O(y))|0\rangle\,=\, \theta(x^0-y^0)\,\langle 0 | O(x) O(y)|0\rangle \,-\,
\theta(y^0-x^0)\,\langle 0 | O(y) O(x)|0\rangle\,,
\]
and using the spectral representation \eqref{eq:spWa1PI} for the two Wightman functions on the right hand side,
it immediately follows that,
\[
\langle 0 | T(O(x) O(y))|0\rangle_{\rm 1PI}\,=\,
\int_0^{\infty} ds \, \rho(s)_{\,\rm 1PI}\, \,\Delta^{(0)}(x-y; s) \,,
\label{eq:spDelta}
\]
where $\Delta^{(0)}(x-y; s)$ is the free-theory expression for the time-ordered (Feynman) propagator,
\[ 
\Delta^{(0)}(x-y; m^2) \,=\, \langle 0| T(\phi(x) \phi(y))|0 \rangle_{\rm free}
 \,=\,
\int\frac{d^4 p}{(2\pi)^4}\,  \frac{i}{p^2-m^2 +i\epsilon} \, e^{-i p\cdot(x-y)}\,,
\label{eq:Del0}
\]
with $m^2=s$.

The time-ordered 1PI Green function in \eqref{eq:spDelta} is easily identified with the $(-i)$ times
the self-energy $\Sigma_X$ of the resonance $X(x)$ that is coupled to the source ${\cal O}[\phi(x)]$.
To see this, consider extending the Lagrangian ${\cal L}[\phi]$
by adding a new real scalar degree of freedom $X$ coupled to ${\cal O}[\phi(x)]$ as follows,
\[
{\cal L}[\phi, X]\,=\, \frac{1}{2}\,\partial_\mu X\partial^\mu X \,-\, \frac{1}{2}\, M_X^2 \, X^2\,-\,
X{\cal O}[\phi]\,+\, {\cal L}[\phi]\,.
\]
The equation of motion for $X(x)$,
\[
(\partial_\mu\partial^\mu + M_x^2) X\,=\, -{\cal O}\,,
\]
implies that the LSZ-amputated 2-point function of the $X$-field is $\langle 0 | T(O(x) O(y))|0\rangle$,
and hence its 1PI part is the self-energy of $X$,
\[
-i\, \Sigma_X (x-y)\,=\, \langle 0 | T(O(x) O(y))|0\rangle_{\rm 1PI}\,=\,
\int_0^{\infty} ds \, \rho(s)_{\,\rm 1PI}\, \,\Delta^{(0)}(x-y; s) \,.
\label{eq:spDeltaSig}
\]

Fourier transform of the time-ordered correlator \eqref{eq:spDeltaSig},
\[
\Sigma_X(p^2)\,=\, i\, \int d^4 x \, e^{i p\cdot x} \, \langle 0 | T(O(x) O(0))|0\rangle
\]
gives a deceptively simple-looking expression,
\[
\Sigma_X(p^2)\,=\,-\,  \int_0^{\infty} ds \,\, \frac{\rho(s)_{\, \rm 1PI}}{p^2-s +i\epsilon}
\,,
\label{eq:disp1}
\]
which is what we have referred to earlier as second K\"all\'en-Lehmann spectral formula.

\medskip 

A sell-known consequence of formula Eq.~\eqref{eq:disp1} is the dispersion relation which implies that 
$\rho(s)_{\,\rm 1PI}= {\rm Im}(\Sigma_X(s)) /\pi$,
if one assumes that $\rho(s) \to 0$ at large $s$, so that one can deform the integration contour in the complex $s$ plane. 
The problem, however, is that the integral $\int ds$ is always divergent in a 4-dimensional theory at $s\to \infty$
and the expression on the right hand side of \eqref{eq:disp1} is ill-defined, see e.g. \cite{Weinberg:1995mt}.

\medskip

If  one now makes an assumption that $\rho(s)_{\,\rm 1PI}$ is a tempered distribution, the formula 
\eqref{eq:disp1} can be recovered after a finite number of subtractions.
Specifically, if the spectral density is tempered, it must grow at $s\to \infty$ no faster than a fixed-order polynomial.
If this polynomial is
of the order $N$,  i.e.
\[
s\to \infty: \qquad \rho(s)_{\rm 1PI}\,\le \, c_0 \,+\,c_1 s\,+\, \ldots\, c_{N} s^{N}\,,
\]
one proceeds to differentiate both sides of the equation \eqref{eq:disp1} $N+1$ times with respect to $p^2$, until the 
integral $\int ds$  on the right hand side of \eqref{eq:disp1} becomes convergent.
This procedure is equivalent to implementing $N+1$ subtractions with unknown integration constants from
the right hand side of \eqref{eq:disp1}, or more precisely,
\[
-i\, \Sigma(p^2)\,=\, P_N(p^2) \,+\, p^{2(N+1)}  \int_0^{\infty} ds \,\, \frac{i}{p^2-s +i\epsilon}\, \frac{\rho(s)_{\rm 1PI}}{s^{N+1}}
\,.
\label{eq:dispN}
\]
Thus we conclude that the assumption that $\rho(s)_{\rm 1PI}$ is a {\it tempered} distribution is equivalent to assuming that the dispersion relation
 \eqref{eq:dispN} can be well-defined by making a {\it finite} number of subtractions. However, there is no reason why the number of subtractions should
 always be finite, for example 
 in non-perturbative settings where $\sqrt{s} \sim m/\lambda$. Strictly localizable theories of Jaffe type with the power of exponential growth 
 $\kappa$ in the regime $0<\kappa<1$ provide a mathematically well-defined local QFT formulation 
 but result in an infinite number of subtractions in \eqref{eq:dispN}. So we see no fundamental reason why the number of subtractions in the
 formally divergent expression \eqref{eq:disp1} should always be finite.

\medskip

The reason why the Fourier transform of the time-ordered correlator \eqref{eq:disp1} turns out to be infinite is that it does not include the 
test functions. 
The corrected expression that includes averaging of the operators $O(x)$ with test functions $g(x)$ reads,
\[
-i\, \hat\Sigma(p^2)\,=\,  \int_0^{\infty} ds \,\, \frac{i}{p^2-s +i\epsilon}\,\, \rho(s)_{\rm 1PI} \cdot |\tilde{g}(\sqrt{s})|^2
\,,
\label{eq:disp1gg}
\]
where the factor $|\tilde{g}(p)|^2$ arises from the Fourier transforms of the smearing (test) functions $g(x)$ and ensures that 
the Feynman propagator does in fact get cut-off at asymptotically large $p^2$.

\medskip

Note that in \eqref{eq:disp1gg} we treat the test function $g(x)$ as the integral part of the definition of the operator
$O_g$  in \eqref{eq:distint} used for computing the correlators. Hence $g(x)$ plays the role of the smearing function 
of the operator. The smeared operator $O_g$ is defined over a vicinity of a point $x$ rather than being sharply defined at the point $x$.

Notice that, strictly speaking, the smearing functions in \eqref{eq:disp1gg}
are only needed in order to 
keep the {\it real part} of the self-energy in \eqref{eq:disp1gg} under control. The {\it imaginary part} of the self-energy 
can be determined with no difficulty already from \eqref{eq:disp1}, by using the fact that the 
spectral density is a real-valued function of $s$ and that
${\rm Im} \frac{1}{p^2-s+i\epsilon} \propto \delta(p^2-s)$. This implies, 
${\rm Im}\, \Sigma (p^2) \,=\, \pi \, \rho(s)_{\rm 1PI}$, and that ${\rm Im}\, \hat\Sigma (p^2)$ is well-defined 
even for $|\tilde{g}|^2 =1$, i.e. without any smearing effects.

\bigskip 

Consider the Higgsplosion process \eqref{eq:1pH} with the initial state $\ket{1^*}$ being a highly virtual off-shell boson
with $p^2=s$. The Higgsplosion rate corresponds to the particle decay width which is proportional to the imaginary part of the 
self-energy,
\[
{\rm Im}\, \hat\Sigma_n (p^2)\,=\,-\, m\,\Gamma_n(p^2)\, \sim\,  |\tilde{g} (\sqrt{p^2})|^2 \,\, {\cal R}_n^{\rm semicl}(\sqrt{p^2})\,,
\label{eq:SigRnhg}
\]
where ${\cal R}_n^{\rm semicl}(\sqrt{p^2})$ is the semiclassical prediction for the Higgsplosion rate, which becomes exponentially large
above a certain energy scale $p^2\ge E_*^2$. As in \eqref{eq:ggRnhg} and \eqref{eq:disp1gg} we have included on the right hand side of 
\eqref{eq:SigRnhg} the smearing effect of the test functions.

The self-energy contribution in \eqref{eq:SigRnhg} can now be resummed to obtain the full Dyson propagator,
\[
\tilde\Delta(p) \,=\, \frac{i }{p^2-m^2\,-\, \hat\Sigma(p^2) +i\epsilon}
\,.
\label{eq:propUVd2}
\]

We can always choose the smearing functions on the right hand side of \eqref{eq:SigRnhg} such that they allow the imaginary part of 
self-energy to become
greater than $p^2$, at the Higgsplosion scale $p_*^2$ i.e. 
$|\hat\Sigma(p_*^2)| \gtrsim p_*^2$ before they cut-off the exponential growth of ${\rm Im}\, \Sigma$ at asymptotically large momenta 
$\gg p_*^2$.
In this case the expression \eqref{eq:propUVd2} for the Dyson propagator becomes exponentially small at the Higgsplosion scale
as the result of the large self-energy contribution in the denominator.
We conclude that the fall off of the propagator at $p^2$ above the Higgsplosion scale (and before the smearing functions cut-off the self-energy
at asymptotically high momenta) is entirely consistent with the phenomenon of Higgspersion 
of the resummed Dyson propagator proposed in \cite{Khoze:2017tjt,Khoze:2017lft}.

\medskip

We would like to add in conclusion that the Dyson-resummed form of the propagator in \eqref{eq:propUVd2},
that is central to the Higgspersion mechanism of \cite{Khoze:2017tjt,Khoze:2017lft},
can also be 
intuitively understood as a result of summing over contributions from multiple saddle-point solutions to the 2-point function,
such as those shown in Fig.~\ref{fig:two}.
These more complicated saddle-points correspond to multi-centre solutions with multiple singularities. 
The logic is similar to the `premature unitarization' approach used in instanton-based
semiclassical calculations in Refs.~\cite{Zakharov:1990xt,Maggiore:1991fc,Maggiore:1991vi,Veneziano:1992rp}. 
In the premature unitarization model,
the total semiclassical amplitude was obtained by summing over general instanton-anti-instanton chains, with the result for the
cross-section being given by summing the geometric progression -- similar in form to the Dyson propagator in \eqref{eq:propUVd2} 
where $\hat\Sigma(p^2)$ comes from the single simple saddle-point solution. The result is that the 
overall effect = the sum of the geometric progression is suppressed at and above  the Higgsplosion scale.

\newpage

%%%%%%%%%%%%%%%%%%%

\section{Conclusions}
\label{sec:4}
%%%%%%%%%%%%%%%%%%%

If Higgsplosion can be realised in the Standard Model, its consequences for particle theory would be astounding.
Higgsplosion would result in an exponential suppression of quantum fluctuations beyond the Higgsplosion energy scale and have observable
consequences at future high-energy colliders and in 
cosmology~\cite{Khoze:2017lft,Jaeckel:2014lya,Gainer:2017jkp,Khoze:2017uga,Khoze:2018bwa}.

\medskip

Production of large numbers of particles in scattering processes at very high energies was studied in great detail
in the classic papers~\cite{Cornwall:1990hh,Goldberg:1990qk,Brown:1992ay,Argyres:1992np,Voloshin:1992rr,
Voloshin:1992nu,Libanov:1994ug}, and more recently in 
\cite{Khoze:2014kka}.
These papers largely relied on calculations in perturbation theory, which in the regime of interest for Higgsplosion, 
$n \gtrsim 1/\lambda \gg 1$,
is strongly coupled and calls for a robust non-perturbative formalism. 
Semiclassical methods~\cite{Gorsky:1993ix,Son:1995wz, Libanov:1997nt}
provide a way to achieve this.

At present, Higgsplosion remains a conjecture based on the application of the semiclassical approach  
of \cite{Son:1995wz} to scalar QFT models of the type  \eqref{eq:ssb} in the non-relativistic large-$n$ steepest descent limit 
in the calculations in~\cite{Khoze:2018kkz,Khoze:2017ifq}.\footnote{The semiclassical technique used 
in~\cite{Khoze:2018kkz,Khoze:2017ifq} is reliant on using QFT (i.e. a system with an infinite number of degrees of freedom) 
in not less than $2+1$ dimensions and spontaneous symmetry breaking. For example, it is known that in a finite dimensional quantum
mechanics the analogues of high-multiplicity amplitudes are exponentially suppressed~\cite{Bachas:1991fd,Jaeckel:2018ipq}.}

\medskip

In the semiclassical limit a theory with Higgsplosion results in an exponentially growing expression for the spectral density distribution 
function $\rho(s)$. This implies that the spectral density cannot be a tempered distribution. This is a trivial statement, it relies solely on the 
definition of tempered distributions -- which are those that grow at large $\sqrt{s}$ at most as a polynomial of $\sqrt{s}$. 
Hence any exponentially 
growing distributions are not tempered. But it is well-established since the work of Jaffe \cite{Jaffe:1967nb} 
that local quantum field theory does not require the assumption of temperedness.

The main purpose of this note was to explain that the semiclassical Higgsplosion is not inherently problematic or inconsistent with the local
QFT framework, contrary to what was implied in the recent articles \cite{Belyaev:2018mtd,Monin:2018cbi}.
Restrictions imposed by the requirement that quantum fields are strictly localizable, in fact, allows the matrix elements to grow faster than any 
polynomial; the admissible distributions need not be tempered.
The upper bound on the growth of momentum space distributions is a linear exponential~\cite{Jaffe:1967nb}, 
and hence the semiclassical expression 
for the spectral density in a theory with Higgsplosion is admissible and consistent with the requirement of strictly localizable fields.
Such strictly local field theories were formulated in precise mathematical form in~\cite{Jaffe:1967nb}.
It was found
that key results in QFT such as the connection between spin and statistics, the existence of CPT symmetry, crossing symmetry, unitarity 
and dispersion relations (allowing for infinite number of subtractions), can be derived and continue to hold in this framework 
without assuming (restricting to) tempered fields.

Other examples of models with exponentially growing spectral density have been studied in the literature. Perhaps the simplest example is
that of the exponential operator of the free field, ${\cal O}(x) = e^{j \phi(x)}$. Its spectral density was discussed and computed in
\cite{Jaffe:1967nb,Lehmann:1971gq,Keltner:2015xda}. The result quoted in \cite{Keltner:2015xda} is,
\[
\rho(s)\,=\, e^{\frac{3j^{2/3}\sqrt{s}^{2/3}}{2^{4/3}\pi^{2/3}}}\, \frac{1}{\sqrt{3}}\left(\frac{2j^4}{s^4\pi}\right)\, \sim e^{c\,(\sqrt{s})^{2/3}} \,.
\]
In an interacting QFT which has a UV fixed point, the high-energy behaviour of the spectral density
of generic operators localised in a volume $V$ was estimated in \cite{Aharony:1998tt} to have the form,
\[
\rho(s) \,\sim \, e^{\,c V^{\frac{1}{d}}(\sqrt{s})^{\frac{d-1}{d}}} \,,
\label{eq:AB}
\]
where $d$ is the number of spacetime dimensions. 

In a gravitational theory in $d$ asymptotically flat space-time dimensions, the spectral
density is dominated by black hole states~\cite{Aharony:1998tt} with $\rho(s) \,\sim \, e^{\,c (\sqrt{s}/M_{\rm Pl})(\sqrt{s})^{\frac{d-2}{d-3}}}$
which corresponds to a non-localizable distribution with $\kappa >1$. 
Galileon field theories \cite{Luty:2003vm,Nicolis:2008in} and Little string theories \cite{Kapustin:1999ci} 
are also characterised by exponentially growing spectral densities 
and fall in the class 
with $\kappa >1$  \cite{Keltner:2015xda}.

\medskip

We note that
if one wishes to relax the requirement of strict localizability,
Higgsplosion will continue to work also with
the non-localizable QFT framework studied in Refs.~\cite{Meiman,Efimov:1967pjn,Iofa:1969fj,Iofa:1969ex,Steinmann:1970cm}.
Only tempered distributions are incompatible with the semiclassical Higgsplosion, 
as, by their very definition, they cannot allow for any form of exponential growth of the spectral density.

%%%%%%%%%%%%%%%%%%%%%%%%%%%%%%%%
%%%%%%%%%%%%%%%%%%%%%%%%%%%%%%%%%%%%%%%%%%%%%%%%%%%%

\bigskip

\section*{Acknowledgements}
We would like to thank J. Jaeckel for useful discussions.

%%%%%%%%%%%%%%%%%%%%%%%%%%%%%

\bibliographystyle{JHEP}
\bibliography{references}

\end{document}